\documentclass[twocolumn,showpacs,preprintnumbers,amsmath,amssymb, showkeys]{revtex4}

\usepackage{graphicx}
\usepackage{dcolumn}
\usepackage{bm}
\usepackage{epsf}

\begin{document}


\title{Charge and spin configurations in the coupled quantum dots with Coulomb correlations induced
by tunneling current}

\author{P.\,I.\,Arseev}
 \altaffiliation{ars@lpi.ru}
\author{N.\,S.\,Maslova}%
 \email{spm@spmlab.phys.msu.ru}
\author{V.\,N.\,Mantsevich}
 \altaffiliation{vmantsev@spmlab.phys.msu.ru}
\affiliation{%
 P.N. Lebedev Physical institute of RAS, 119991, Moscow, Russia\\~\\
 Moscow State University, Department of  Physics,
119991 Moscow, Russia
}%

\date{\today }

\begin{abstract}
We investigated the peculiarities of non-equilibrium charge states
and spin configurations in the system of two strongly coupled
quantum dots (QDs) weakly connected to the electrodes in the
presence of Coulomb correlations. We analyzed the modification of
non-equilibrium charge states and different spin configurations of
the system in a wide range of applied bias voltage and revealed well
pronounced ranges of system parameters where negative tunneling
conductivity appears due to the Coulomb correlations.
\end{abstract}

\maketitle

\section{Introduction}

Electron tunneling through the system of coupled quantum dots in the
presence of strong Coulomb correlations seems to be one of the most
interesting problems in the solid state physics. The present day
experimental technique gives possibility to produce single QDs with
a given set of parameters and to create coupled QDs with different
spatial geometry
\cite{Vamivakas},\cite{Stinaff},\cite{Elzerman},\cite{Munoz-Matutano}.
Well known vertically aligned geometry
\cite{Vamivakas},\cite{Stinaff},\cite{Elzerman} gives an opportunity
to analyze non-stationary effects in various charge and spin
configurations formation in the small size structures both
theoretically \cite{Kikoin_1} and experimentally \cite{Elzerman}.
Lateral QDs are also actively studied experimentally during the last
several years \cite{Munoz-Matutano}, but due to the technological
problems they are mostly analyzed theoretically
\cite{Peng},\cite{Szafran}. Thereby the main effort in the physics
of QDs is devoted to the investigation of non-equilibrium charge
states and different spin configurations due to the electrons
tunneling
\cite{Goldin},\cite{Kikoin},\cite{Kikoin_1},\cite{Orellana} through
the system of coupled QDs in the presence of strong Coulomb
interaction. One of the most intensively studied problems in this
field is tunneling through the single QD \cite{Paaske_1},
\cite{Paaske_2}, \cite{Kaminski} and interacting QDs
\cite{Orellana},
\cite{Lopez},\cite{Goldin},\cite{Kikoin},\cite{Kikoin_1} in the
Kondo regime, which reveals rich physics for small bias voltage
compared to the tunneling rates. Theoretical analysis of this
problem usually deals with the Keldysh non-equilibrium
Green-function formalism \cite{Goldin},\cite{Paaske},
renormalization-group theory \cite{Kikoin} or specific approach
suggested by Coleman
\cite{Coleman},\cite{Coleman_1},\cite{Paaske},\cite{Orellana}.

One of the most interesting results was obtained for the system of
double QDs with on site Coulomb repulsion in both of them
\cite{Orellana}. The authors demonstrated that due to the presence
of Coulomb correlations QDs can have a bistable behavior in the
Kondo regime at zero bias voltage.

Charge redistribution between different spin configurations in the
system of two interacting QDs in the Kondo regime was regarded in
\cite{Kikoin_1}. The authors considered the situation when the
detuning between the energy levels in the QDs exceeds the dots
coupling and on-site Coulomb repulsion is present only in a single
dot. A new mechanism which leads to the transition from the singlet
state in a weak coupling regime to a triplet state in a strong
coupling regime was proposed. The authors demonstrated that
interaction with continuous spectrum at zero bias modifies the
energy of the singlet and triplet states and the situation when
triplet state energy is lower than singlet one becomes possible. So
careful analysis of tunneling processes through the system of
interacting QDs in the Kondo regime reveals exciting physical
phenomena.

In the present paper we consider electron tunneling through the
coupled QDs in the regime when applied bias can be tuned in a wide
range and the on-site Coulomb repulsion can be comparable to the
other system parameters. We analyze all charge and spin
configurations in the system of two strongly coupled quantum dots
(QDs) weakly connected to the electrodes in the presence of Coulomb
correlations in a wide range of applied bias in terms of pseudo
operators with constraint
\cite{Coleman},\cite{Coleman_1},\cite{Barnes},\cite{Barnes_1},\cite{Bickers},\cite{Wingreen}.
For large values of applied bias Kondo effect is not essential so we
neglect any correlations between electron states in the QDs and in
the leads. This approximation allows to describe correctly non
equilibrium occupation of any single- and multi-electron state due
to  the tunneling processes.

We revealed the presence of negative tunneling conductivity in
certain ranges of the applied bias voltage and analyzed the multiple
charge redistribution between the two electron states with different
spin configurations (singlet state and triplet state) as a function
of the applied bias voltage.

\section{Theoretical model}

We consider a system of coupled QDs with the single particle levels
$\widetilde{\varepsilon}_{1}$ è $\widetilde{\varepsilon}_{2}$
connected to the two leads. Such system can be described by means of
two-impurities Anderson Hamiltonian where the impurities are the
QD's \cite{Anderson},\cite{Aguado},\cite{Busser}. The Hamiltonian
can be written as:

\begin{eqnarray}
\Hat{H}&=&\sum_{\sigma}a_{1\sigma}^{+}a_{1\sigma}\widetilde{\varepsilon}_{1}+
\sum_{\sigma}a_{2\sigma}^{+}a_{2\sigma}\widetilde{\varepsilon}_{2}+
U_1\widehat{n}_{1\sigma}\widehat{n}_{1-\sigma}+\nonumber\\&+&U_2\widehat{n}_{2\sigma}\widehat{n}_{2-\sigma}+
\sum_{\sigma}T(a_{1\sigma}^{+}a_{2\sigma}+a_{2\sigma}^{+}a_{1\sigma})
\end{eqnarray}

The operator $a_{i\sigma}$ creates an electron in the dot $i$ with
spin $\sigma$, $\widetilde{\varepsilon}_{i}$ is the energy of the
single electron level in the dot $i$ and $T$ is the inter-dot
tunneling coupling, $n_{i\sigma}=a_{i\sigma}^{+}a_{i\sigma}$ and
$U_{1,2}$ is the on-site Coulomb repulsion of localized electrons.
We'll consider for simplicity the situation when resonant tunneling
between the QDs takes place and Coulomb repulsion is the same in the
first and second QDs, consequently
$\widetilde{\varepsilon}_{1}=\widetilde{\varepsilon}_{2}=\varepsilon_{0}$
and $U_1=U_2=U$. Without the interaction with the leads all energies
of single- and multi-electron states are well known:

One electron in the system: two single electron states with energies
$\varepsilon_i=\varepsilon_0\pm T$ and wave function

\begin{eqnarray}
\psi_i=\frac{1}{\sqrt{2}}\cdot(|0\uparrow\rangle|00\rangle\pm|00\rangle|0\uparrow\rangle)
\end{eqnarray}

Two electrons in the system: two states with the same spin
$\sigma\sigma$ and $-\sigma-\sigma$ (triplet states has the spin
projection $S_{Z}=\pm1$) with energies $2\varepsilon_{0}$ and four
two-electron states with the opposite spins $\sigma-\sigma$ and
different configurations with energies $E_{IIj}^{\sigma\sigma^{'}}$:
$2\varepsilon_{0}$; $2\varepsilon_{0}+U$ and
$2\varepsilon_{0}\pm\frac{U}{2}+\sqrt{\frac{U^{2}}{4}+4T^{2}}$. Wave
functions have the form:

\begin{eqnarray}
\psi_{j}^{\sigma-\sigma}&=&\alpha\cdot|\uparrow\downarrow\rangle|00\rangle-\beta\cdot|\downarrow0\rangle|0\uparrow\rangle+\nonumber\\&+&
\gamma\cdot|0\uparrow\rangle|\downarrow0\rangle+\delta\cdot|00\rangle|\uparrow\downarrow\rangle\nonumber\\
\end{eqnarray}

Three electrons in the system: two three-electron states with
energies $E_{III}^{m\sigma}=3\varepsilon_0+U\pm T$ and wave function

\begin{eqnarray}
\psi_{m\sigma}&=&\frac{1}{\sqrt{2}}\cdot|\downarrow0\rangle|\downarrow0\rangle\cdot(|0\uparrow\rangle|00\rangle\pm|00\rangle|0\uparrow\rangle)\nonumber\\
m&=&\pm1
\end{eqnarray}

Four electrons in the system: one four-electron state with energy
$E_{IVl}=4\varepsilon_0+2U$ and wave function

\begin{eqnarray}
\psi_{l}=|\uparrow\downarrow\rangle|\uparrow\downarrow\rangle
\end{eqnarray}

If coupled QDs are connected with the leads of the tunneling contact
the number of electrons in the dots changes due to the tunneling
processes. Transitions between the states with different number of
electrons in the two interacting QDs can be analyzed in terms of
pseudo-particle operators with constraint on the physical states
(the number of pseudo-particles). Consequently, the electron
operator $c_{\sigma}^{+}$ can be written in terms of pseudo-particle
operators as:

\begin{eqnarray}
c_{\sigma}^{+}&=&\sum_{i}f_{\sigma
i}^{+}b+\sum_{j,i}d_{j}^{+\sigma-\sigma}f_{i-\sigma}+\sum_{j,i}d_{j}^{+\sigma\sigma}f_{i\sigma}+\nonumber\\&+&\sum_{m,j}\psi_{m-\sigma}^{+}d_{j}^{\sigma-\sigma}+\sum_{m,j}\psi_{m\sigma}^{+}d_{j}^{-\sigma-\sigma}+\sum_{l}\varphi_{l}^{+}\psi_{m\sigma}\nonumber\\
\end{eqnarray}

where $f_{\sigma}^{+}(f_{\sigma})$ and
$\psi_{\sigma}^{+}(\psi_{\sigma})$- are pseudo-fermion
creation(annihilation) operators for the electronic states with one
and three electrons correspondingly. $b^{+}(b)$,
$d_{\sigma}^{+}(d_{\sigma})$ and $\varphi^{+}(\varphi)$- are slave
boson operators, which correspond to the states without any
electrons, with two electrons or four electrons. Operators
$\psi_{m-\sigma}^{+}$- describe system configuration with two spin
up electrons $\sigma$ and one spin down electron $-\sigma$ in the
symmetric and asymmetric states.

The constraint on the space of the possible system states have to be
taken into account:

\begin{eqnarray}
\widehat{n}_{b}+\sum_{i\sigma}\widehat{n}_{fi\sigma}+\sum_{j\sigma\sigma^{'}}\widehat{n}_{dj}^{\sigma\sigma^{'}}+\sum_{m\sigma}\widehat{n}_{\psi
m\sigma}+\widehat{n}_{\varphi}=1 \label{limit}
\end{eqnarray}

Condition (\ref{limit}) means that the appearance  of any two
pseudo-particles in the system simultaneously is impossible.

Electron filling numbers in the coupled QDs can be expressed in the
terms of the pseudo-particles filling numbers:

\begin{eqnarray}
\widehat{n}_{\sigma}^{el}&=&c_{\sigma}^{+}c_{\sigma}=\sum_{i}\widehat{n}_{fi\sigma}+\sum_{ij}\widehat{n}_{dj}^{\sigma-\sigma}+\sum_{ij}\widehat{n}_{dj}^{\sigma\sigma}+\nonumber\\&+&\sum_{mj}\widehat{n}_{\psi
m-\sigma}+\sum_{mj}\widehat{n}_{\psi
m\sigma}+\sum_{m}\widehat{n}_{\varphi l}
\end{eqnarray}

Consequently, the Hamiltonian of the system can be written in the
terms of the pseudo-particle operators:

\begin{eqnarray}
\Hat{H}&=&\Hat{H_{0}}+\Hat{H}_{tun}\\
\Hat{H_{0}}&=&\sum_{i\sigma}\varepsilon_{i}f_{i\sigma}^{+}f_{i\sigma}+\sum_{j\sigma\sigma^{'}}E_{IIj}^{\sigma\sigma^{'}}d_{j}^{+\sigma\sigma^{'}}d_{j}^{\sigma\sigma^{'}}+\nonumber\\&+&\sum_{m\sigma}E_{III}^{m\sigma}\psi_{m\sigma}^{+}\psi_{m\sigma}+E_{IVl}\varphi_{l\sigma}^{+}\varphi_{l\sigma}+\nonumber\\&+&\sum_{k\sigma}\varepsilon_{k\sigma}c_{k\sigma}^{+}c_{k\sigma}+\sum_{p\sigma}(\varepsilon_{p\sigma}-eV)c_{p\sigma}^{+}c_{p\sigma}\nonumber\\
\Hat{H}_{tun}&=&\sum_{k\sigma}T_{k}(c_{k\sigma}^{+}c_{\sigma}+c_{\sigma}^{+}c_{k\sigma})+(k\leftrightarrow
p)\nonumber\
\end{eqnarray}

where $\varepsilon_i$, $E_{IIj}^{\sigma\sigma^{'}}$,
$E_{III}^{m\sigma}$ and $E_{IVl}$-are the energies of the single-,
double-, triple- and quadri-electron states.
$\varepsilon_{k(p)\sigma}$-is the energy of the conduction electrons
in the states $k$ and $p$ correspondingly.
$c_{k(p)\sigma}^{+}/c_{k(p)\sigma}$ are the creation(annihilation)
operators in the leads of the tunneling contact. $T_{k(p)}$-are the
tunneling amplitudes, which we assume to be independent on momentum
and spin. Indexes $k(p)$ mean only that tunneling takes place from
the system of coupled QDs to the conduction electrons in the states
$k$ and $p$ correspondingly.

Bilinear combinations of pseudo-particle operators are closely
connected with the density matrix elements. So, similar expressions
can be obtained from equations for the density matrix evolution but
method based on the pseudo particle operators is more compact and
convenient. The tunneling current through the proposed system
written in terms of the pseudo-particle operators has the form:

\begin{eqnarray}
\widehat{I}_{k\sigma}&=&\sum_{k}\frac{\partial
\widehat{n}_{k}}{\partial t}=i \left[\sum_{ik}T_{k}
c_{k\sigma}f_{i\sigma}^{+}b+\sum_{ijk}T_{k}
c_{k\sigma}d_{j}^{+\sigma-\sigma}f_{i-\sigma}+\right.\nonumber\\&+&\sum_{ijk}T_{k}
c_{k\sigma}d_{j}^{+\sigma\sigma}f_{i\sigma}+
\sum_{mjk}T_{k}c_{k\sigma}\psi_{m-\sigma}^{+}d_{j}^{\sigma-\sigma}+\nonumber\\
&+&\left.
\sum_{mjk}T_{k}c_{k\sigma}\psi_{m\sigma}^{+}d_{j}^{-\sigma-\sigma}+\sum_{mk}T_{k}c_{k\sigma}\varphi_{l}^{+}\psi_{m\sigma}-h.c.\right]
\end{eqnarray}

We set $\hbar=1$ and neglect changes in the electron spectrum and
local density of states in the tunneling contact leads, caused by
the tunneling current. Therefore equations of motion together with
the constraint on the space of the possible system states
(pseudo-particles number) (\ref{limit}) give the following
equations:
\begin{widetext}
\begin{eqnarray}
Im\sum_{ik}T_{k}\cdot \langle
c_{k\sigma}f_{i\sigma}^{+}b\rangle&=&\Gamma_{k}\sum_{i}[(1-n_{k\sigma}(\varepsilon_i))\cdot
n_{fi\sigma}-n_{k\sigma}(\varepsilon_{i})\cdot n_{b}]\nonumber\\
Im\sum_{ijk}T_{k}\cdot \langle
c_{k\sigma}d_{j}^{+\sigma-\sigma}f_{i-\sigma}\rangle&=&\Gamma_{k}\sum_{ij}[(1-n_{k\sigma}(E_{IIj}^{\sigma-\sigma}-\varepsilon_{i-\sigma}))\cdot
n_{dj}^{\sigma-\sigma}-n_{k\sigma}(E_{IIj}^{\sigma-\sigma}-\varepsilon_{i-\sigma})\cdot
n_{fi-\sigma}]\nonumber\\
Im\sum_{ijk}T_{k}\cdot \langle
c_{k\sigma}d_{j}^{+\sigma\sigma}f_{i\sigma}\rangle&=&\Gamma_{k}\sum_{ij}[(1-n_{k\sigma}(E_{IIj}^{\sigma\sigma}-\varepsilon_{i\sigma}))\cdot
n_{dj}^{\sigma\sigma}-n_{k\sigma}(E_{IIj}^{\sigma\sigma}-\varepsilon_{i\sigma})\cdot
n_{fi\sigma}]\nonumber\\
Im\sum_{mjk}T_{k}\cdot \langle
c_{k\sigma}\psi_{m-\sigma}^{+}d_{j}^{\sigma-\sigma}\rangle&=&\Gamma_{k}\sum_{mj}[(1-n_{k\sigma}(E_{III}^{m-\sigma}-E_{IIj}^{\sigma-\sigma}))\cdot
n_{\psi
m-\sigma}-n_{k\sigma}(E_{III}^{m-\sigma}-E_{IIj}^{\sigma-\sigma})\cdot
n_{dj}^{\sigma-\sigma}]\nonumber\\
Im\sum_{mjk}T_{k}\cdot \langle
c_{k\sigma}\psi_{m\sigma}^{+}d_{j}^{-\sigma-\sigma}\rangle&=&\Gamma_{k}\sum_{mj}[(1-n_{k\sigma}(E_{III}^{m\sigma}-E_{IIj}^{-\sigma-\sigma}))\cdot
n_{\psi
m\sigma}-n_{k\sigma}(E_{III}^{m\sigma}-E_{IIj}^{-\sigma-\sigma})\cdot
n_{dj}^{-\sigma-\sigma}]\nonumber\\
Im\sum_{mk}T_{k}\cdot \langle
c_{k\sigma}\varphi_{l}^{+}\psi_{m\sigma}\rangle&=&\Gamma_{k}\sum_{m}[(1-n_{k\sigma}(E_{IVl}-E_{III}^{m\sigma}))\cdot
n_{\varphi}-n_{k\sigma}(E_{IVl}-E_{III}^{m\sigma})\cdot n_{\psi
m\sigma}] \label{tunneling_current}
\end{eqnarray}
\end{widetext}

Tunneling current $I_{k\sigma}$ is determined by the sum of the
right hand parts of the equations (\ref{tunneling_current}).

Stationary system of equations can be obtained for the pseudo
particle filling numbers  $n_{fi}$, $n_{dj}^{\sigma-\sigma}$,
$n_{d}^{\sigma\sigma}$, $n_{\psi m}$ and $n_{\varphi}$:

\begin{widetext}
\begin{eqnarray}
0=\frac{\partial n_{\varphi}}{\partial
t}&=&-\Gamma_{k}\sum_{m\sigma}[-n_{\psi m\sigma}\cdot
n_{k\sigma}(E_{IVl}-E_{III}^{m\sigma}-\varepsilon_{k})+n_{\varphi}\cdot(1-n_{k\sigma}(E_{IVl}-E_{III}^{m\sigma}-\varepsilon_{k}))]+(k\leftrightarrow p)\nonumber\\
0=\frac{\partial n_{\psi m\sigma}}{\partial
t}&=&-\Gamma_{k}\sum_{j}[n_{\psi m\sigma}\cdot
(1-n_{k-\sigma}(E_{III}^{m\sigma}-E_{IIj}^{\sigma-\sigma}-\varepsilon_{k}))-n_{k-\sigma}(E_{III}^{m\sigma}-E_{IIj}^{\sigma-\sigma}-\varepsilon_{k})\cdot
n_{dj}^{\sigma-\sigma}]-\nonumber\\
&-&\Gamma_{k}\sum_{j}[(1-n_{k\sigma}(E_{III}^{m\sigma}-E_{IIj}^{-\sigma-\sigma}-\varepsilon_{k}))\cdot
n_{\psi m\sigma}-n_{dj}^{-\sigma-\sigma}\cdot
n_{k\sigma}(E_{III}^{m\sigma}-E_{IIj}^{-\sigma-\sigma}-\varepsilon_{k})]-\nonumber\\
&-&\Gamma_{k}[-(1-n_{k\sigma}(E_{IVl}-E_{III}^{m\sigma}-\varepsilon_{k}))\cdot
n_{\varphi}+n_{\psi m\sigma}\cdot n_{k\sigma}(E_{IVl}-E_{III}^{m\sigma}-\varepsilon_{k})]+(k\leftrightarrow p)\nonumber\\
0=\frac{\partial n_{dj}^{\sigma\sigma}}{\partial
t}&=&-\Gamma_{k}\sum_{i}[(1-n_{k\sigma}(E_{IIj}^{\sigma\sigma}-\varepsilon_{i}-\varepsilon_{k}))\cdot
n_{dj}^{\sigma\sigma}-n_{k\sigma}(E_{IIj}^{\sigma\sigma}-\varepsilon_{i}-\varepsilon_{k})\cdot n_{fi\sigma}]-\nonumber\\
&-&\Gamma_{k}\sum_{m}[n_{k-\sigma}(E_{III}^{m-\sigma}-E_{IIj}^{\sigma\sigma}-\varepsilon_{k})\cdot
n_{dj}^{\sigma\sigma}-(1-n_{k-\sigma}(E_{III}^{m-\sigma}-E_{IIj}^{\sigma\sigma}-\varepsilon_{k}))\cdot
n_{\psi
m-\sigma}]+(k\leftrightarrow p)\nonumber\\
0=\frac{\partial n_{dj}^{\sigma-\sigma}}{\partial
t}&=&-\Gamma_{k}\sum_{i\sigma}[(1-n_{k-\sigma}(E_{IIj}^{\sigma-\sigma}-\varepsilon_{i}-\varepsilon_{k}))\cdot
n_{dj}^{\sigma-\sigma}-n_{k-\sigma}(E_{IIj}^{\sigma-\sigma}-\varepsilon_{i}-\varepsilon_{k})\cdot n_{fi\sigma}]-\nonumber\\
&-&\Gamma_{k}\sum_{m\sigma}[n_{k\sigma}(E_{III}^{m\sigma}-E_{IIj}^{\sigma-\sigma}-\varepsilon_{k})\cdot
n_{dj}^{\sigma-\sigma}-(1-n_{k\sigma}(E_{III}^{m\sigma}-E_{IIj}^{\sigma-\sigma}-\varepsilon_{k}))\cdot
n_{\psi
m-\sigma}]+(k\leftrightarrow p)\nonumber\\
0=\frac{\partial n_{fi\sigma}}{\partial
t}&=&\Gamma_{k}[n_{k\sigma}(\varepsilon_i-\varepsilon_k)\cdot
n_{b}-(1-n_{k\sigma}(\varepsilon_i-\varepsilon_k))\cdot n_{fi\sigma}]+\nonumber\\
&+&\Gamma_{k}\sum_{j\sigma}[(1-n_{k-\sigma}(E_{IIj}^{\sigma-\sigma}-\varepsilon_i-\varepsilon_k))\cdot
n_{dj}^{\sigma-\sigma}-n_{k-\sigma}(E_{IIj}^{\sigma-\sigma}-\varepsilon_i-\varepsilon_k)\cdot
n_{fi\sigma}]+\nonumber\\
&+&\Gamma_{k}\sum_{i}[(1-n_{k\sigma}(E_{IIj}^{\sigma\sigma}-\varepsilon_{i}-\varepsilon_{k}))\cdot
n_{dj}^{\sigma\sigma}-n_{k\sigma}(E_{IIj}^{\sigma\sigma}-\varepsilon_{i}-\varepsilon_{k})\cdot
n_{fi\sigma}]+(k\leftrightarrow p) \label{system}
\end{eqnarray}
\end{widetext}

In these equations we neglect the non-diagonal averages of
pseudo-particle operators such as $\langle
f_{\sigma}^{+}bd^{+}f_{\sigma}\rangle$ etc.. These terms are of the
next order in small parameter $\Gamma_{k(p)}/\Delta E$ where $\Delta
E$ is the energy difference between any energy states in the coupled
QDs. We consider the paramagnetic situation, when conditions
$n_{fi\sigma}=n_{fi-\sigma}$, $n_{\psi m\sigma}=n_{\psi m-\sigma}$,
$n_{k\sigma}=n_{k-\sigma}$ and
$n_{dj}^{-\sigma-\sigma}=n_{dj}^{\sigma\sigma}$ are fulfilled.
System of equations (\ref{system}) in the stationary case is the
linear system, which allows to determine pseudo particle filling
numbers, electron filling numbers $n_{el}(eV)$ and tunneling current
$I_{k\sigma}$.

It is necessary to mention that tunneling current through the
proposed system can be also analyzed in usual terms of the electrons
creation/annihilation operators $a_{i\sigma}^{+}/a_{i\sigma}$ and
$c_{k\sigma}^{+}/c_{k\sigma}$ in the localized and continuous
spectrum states correspondingly:

\begin{eqnarray}
I=I_{k\sigma}=I_{k\sigma}=t_{k}(\langle
c_{k\sigma}^{+}a_{1\sigma}\rangle-\langle
a_{1\sigma}^{+}c_{k\sigma}\rangle)
\end{eqnarray}

\begin{figure*} [t]
\includegraphics[width=170mm]{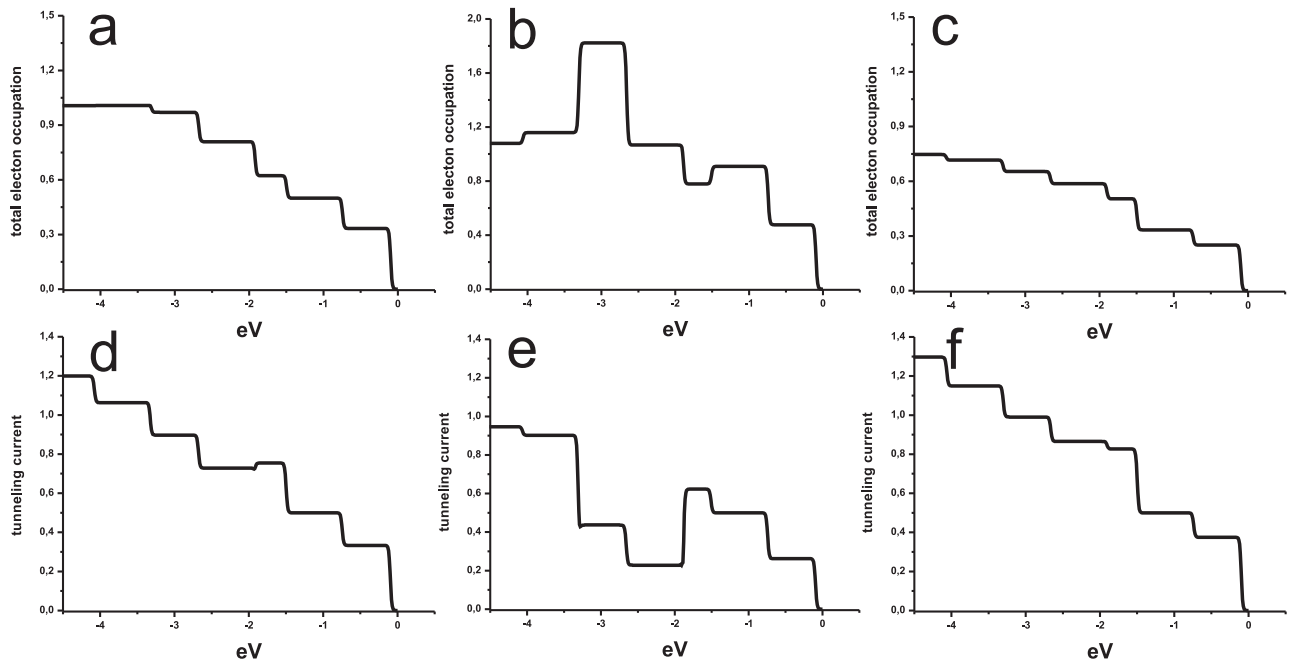}
\caption{Total electron occupation of the QDs a).-c). and tunneling
current d).-f). as a functions of the applied bias voltage for both
single electron energy levels located above the sample Fermi level.
Parameters $\varepsilon/\Gamma=0.8$, $T/\Gamma=0.7$, $U/\Gamma=1.85$
are the same for all figures. a).,d). $\Gamma_k=0.01$,
$\Gamma_p=0.01$; b).,e). $\Gamma_k=0.1$, $\Gamma_p=0.01$; c).,f).
$\Gamma_k=0.01$, $\Gamma_p=0.02$.} \label{figure_1a_1f}
\end{figure*}

By means of Heisenberg equations of motion one can get system of
equations exactly taking into account correlations of electron
filling numbers in localized states in all orders
\cite{Mantsevich},\cite{Mantsevich_2} (for weak tunneling coupling
to the leads). But it is rather tedious to restore the information
about the definite charge and spin configurations with different
number of electrons from all order correlators for initial levels
occupation numbers. So the method based on the pseudo particle
operators is more convenient if we are interested in occupation of
multi-electron states with particular charge and spin configuration.

\begin{figure*} [t]
\includegraphics[width=170mm]{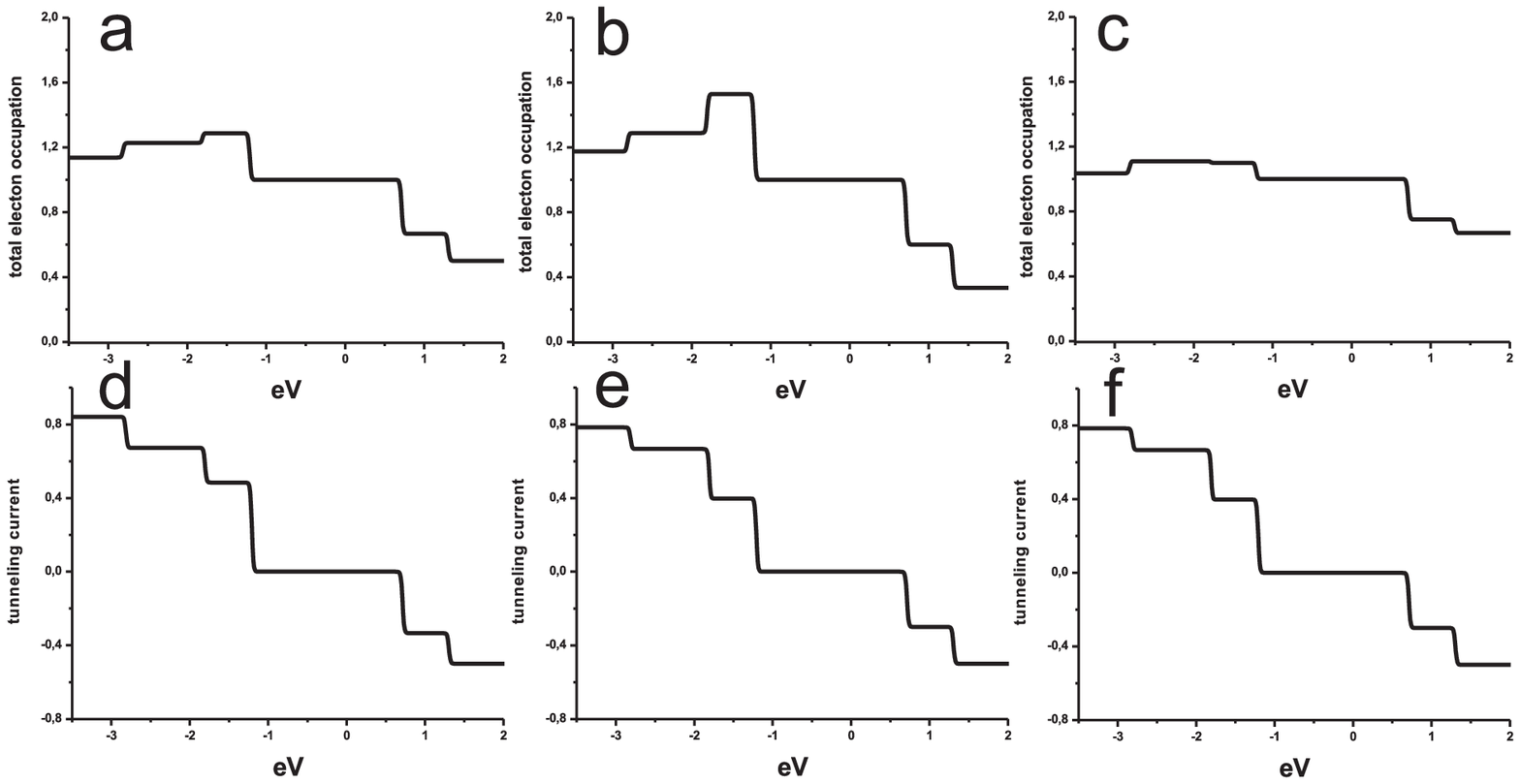}
\caption{The same dependencies as in the Fig.\ref{figure_1a_1f} but
one single electron energy level is located above and the other -
below the sample Fermi level. Parameters $\varepsilon/\Gamma=-0.5$,
$T/\Gamma=0.8$, $U/\Gamma=1.5$ are the same for all figures. a).,d).
$\Gamma_k=0.01$, $\Gamma_p=0.01$; b).,e). $\Gamma_k=0.02$,
$\Gamma_p=0.01$; c).,f). $\Gamma_k=0.01$, $\Gamma_p=0.02$.}
\label{figure_2a_2f}
\end{figure*}

\section{Results and discussion}

\begin{figure*} [t]
\includegraphics[width=170mm]{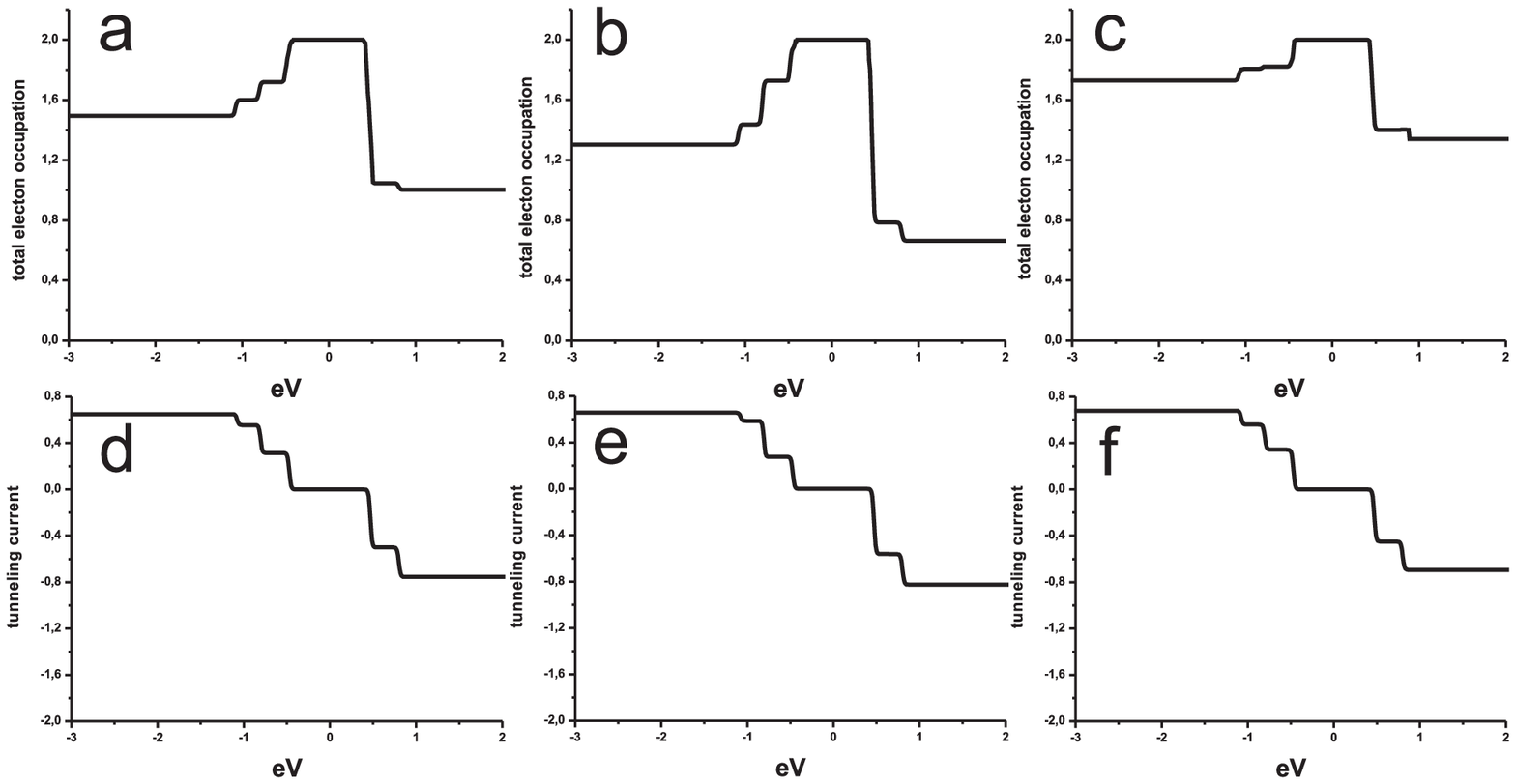}
\caption{The dependencies of total electron occupation of the QDs
a).-c). and tunneling current d).-f). on the applied bias voltage
for both single electron energy levels located below the sample
Fermi level. Parameters $\varepsilon/\Gamma=-0.5$, $T/\Gamma=0.3$,
$U/\Gamma=1.0$ are the same for all cases. a).,d). $\Gamma_k=0.01$,
$\Gamma_p=0.01$; b).,e). $\Gamma_k=0.02$, $\Gamma_p=0.01$; c).,f).
$\Gamma_k=0.01$, $\Gamma_p=0.02$.} \label{figure_3a_3f}
\end{figure*}

The behavior of the total electron occupation of the coupled QDs
$n_{el}(eV)$ and $I-V$ characteristics strongly depends on the
parameters of the tunneling contact: energy levels position, the
value of the Coulomb interaction and the relation between tunneling
rates. The general features of the obtained results is the step-like
$I-V$ characteristics with non-equidistant steps related to the
energies of the different multi-electron states in the QDs and
multiple charge redistribution between the two electron states with
different spin configurations (singlet state and triplet state)
which appears for the particular range of the system parameters and
applied bias voltage.

We first analyze the behavior of the the total electron occupation
of the QDs $n_{el}(eV)$ and $I-V$ characteristics of the considered
system for different single electron levels positions relative to
the sample Fermi level and various tunneling rates to the contact
leads (Fig. \ref{figure_1a_1f}-\ref{figure_3a_3f}). The bias voltage
in our calculations is applied to the sample. Consequently, if both
single electron levels are above(below) the Fermi level, all the
specific features of the total electron occupation and tunneling
current characteristics can be observed at negative(positive) values
of $eV$. In the case when both single electron energy levels are
situated above the sample Fermi level (Fig.\ref{figure_1a_1f}) we
observe the step-like behavior of the total electron occupation both
for the symmetric ($\Gamma_k=\Gamma_p$) and asymmetric
($\Gamma_k<\Gamma_p$) tunneling contact (Fig.\ref{figure_1a_1f}a,c).
The width and height of the steps are determined by the relation
between the system parameters $T$, $\varepsilon$ and $U$ and
$\Gamma=\Gamma_k+\Gamma_p$.

In the particular range of the applied bias when condition
$\Gamma_k>\Gamma_p$ is fulfilled the total electron occupation
demonstrate significant jumps (Fig.\ref{figure_1a_1f}b).

The tunneling current is depicted in  Figures \ref{figure_1a_1f}d-f
as a function of the applied bias (tunneling current amplitudes are
normalized to $2\frac{\Gamma_k\Gamma_p}{\Gamma_k+\Gamma_p}$). In the
asymmetric tunneling contact when condition $\Gamma_k<\Gamma_p$ is
valid (Fig.\ref{figure_1a_1f}f) the tunneling current dependence on
the applied bias has a step-like structure. In the presence of
Coulomb interaction negative tunneling conductivity appears even for
symmetric tunneling contact (Fig.\ref{figure_1a_1f}d). The negative
differential conductivity is strongly pronounced in the asymmetric
tunneling contact when the condition $\Gamma_k>\Gamma_p$ is
fulfilled (Fig.\ref{figure_1a_1f}e).

In a QD without Coulomb interaction the total system occupation can
only increase as applied bias increases and passes single electron
levels. Quite different situation occurs in a system with strong
Coulomb correlations. In the case when one of the single electron
energy levels or both of them are situated below the sample Fermi
level (Fig.\ref{figure_2a_2f}-\ref{figure_3a_3f}) the total electron
occupation of QDs demonstrates pronounced decreasing with increasing
 of the applied bias (Fig. \ref{figure_2a_2f}a-c;
Fig.\ref{figure_3a_3f}a-c). Equilibrium occupation of two-electron
states for zero bias leads to average occupation of one electron per
spin (total occupation equal to 2). But when the increasing bias
reaches the energies of multi-electron excited states, the total
occupation begins to decrease. Using single-electron language we can
say that additional tunneling electrons "push out" electrons from
the states below the Fermi level due to Coulomb repulsion. Or one
can look at this effect as increasing of probability for electrons
to leave the QD due to appearance of several non-elastic channels of
tunneling (accompanied with changing of multi-electron states of the
QD).

The tunneling current as a function of the applied bias for this
case is depicted in Figures
\ref{figure_2a_2f}d-f;\ref{figure_3a_3f}d-f and reveals the
monotonic step-like behavior. In both cases the upper single
electron energy level ($\varepsilon_0+T$) does not appear as a step
in the $I-V$ characteristics.

\begin{figure*} [h]
\includegraphics[width=170mm]{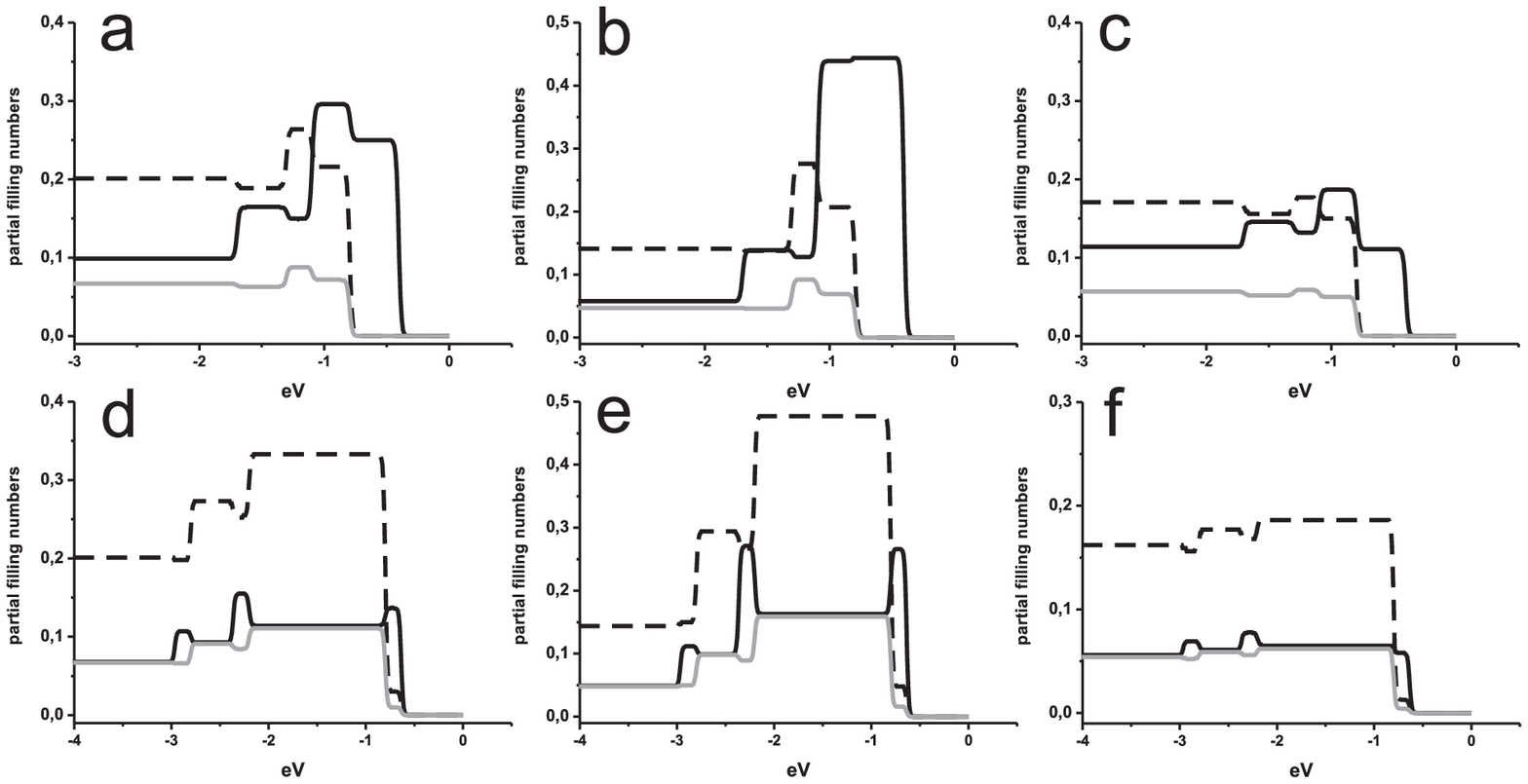}
\caption{Occupation of two electron state for different spin
configurations as a functions of the applied bias voltage for both
single electron levels located above the sample Fermi level. Filling
numbers in the singlet state are shown by the black line, filling
numbers in the triplet state with the fixed projection of the spin
are shown by the grey line, full filling numbers in the triplet
state are shown by the black-dashed line. Parameters
$\varepsilon/\Gamma=0.5$, $T/\Gamma=0.3$ are the same for all
figures. a).,d). $\Gamma_k=0.01$, $\Gamma_p=0.01$; b).,e).
$\Gamma_k=0.02$, $\Gamma_p=0.01$; c).,f). $\Gamma_k=0.01$,
$\Gamma_p=0.02$. a).-c). $U/\Gamma=0.5$; d).-f). $U/\Gamma=2.0$.}
\label{figure_4a_4f}
\end{figure*}

\begin{figure*} [h]
\includegraphics[width=170mm]{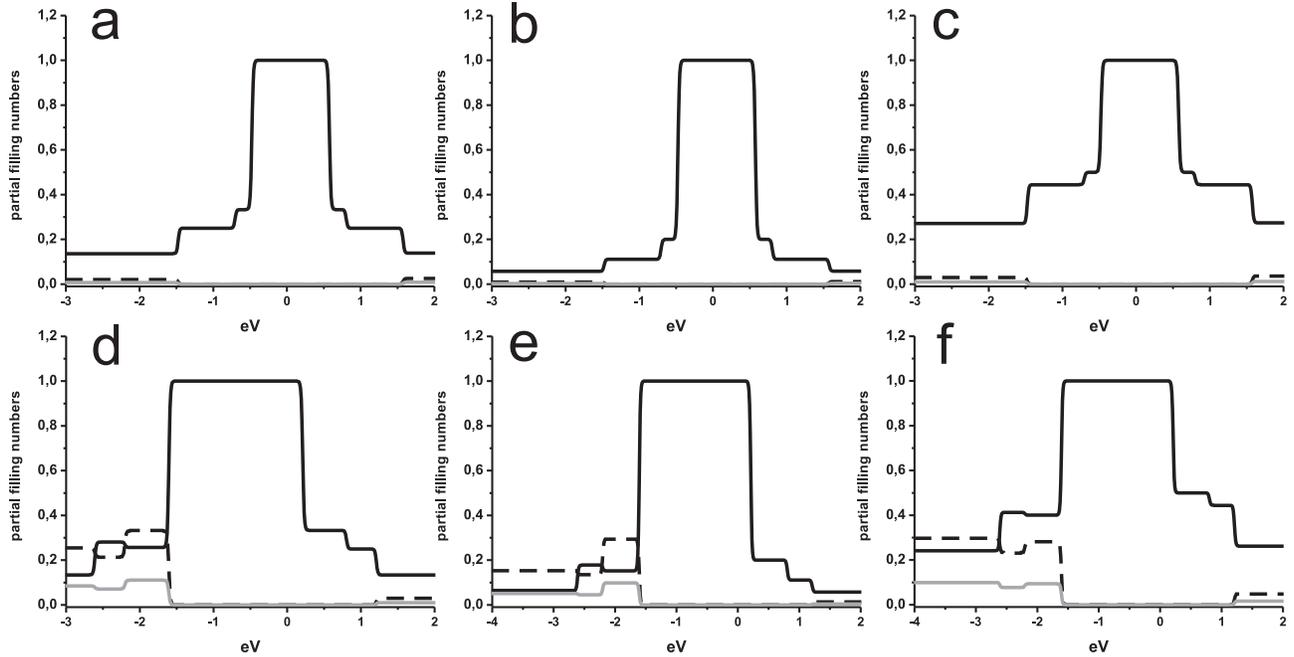}
\caption{The dependence of two electron state filling numbers for
different spin configurations on applied bias voltage when one of
the single electron energy levels is located above and the other-
below the sample Fermi level. Filling numbers in the singlet state
are shown by the black line, filling numbers in the triplet state
with the fixed projection of the spin are shown by the grey line,
full filling numbers in the triplet state are shown by the
black-dashed line. Parameters $\varepsilon/\Gamma=-0.3$,
$T/\Gamma=0.5$ are the same for all figures. a).,d).
$\Gamma_k=0.01$, $\Gamma_p=0.01$; b).,e). $\Gamma_k=0.02$,
$\Gamma_p=0.01$; c).,f). $\Gamma_k=0.01$, $\Gamma_p=0.02$. a).-c).
$U/\Gamma=0.5$; d).-f). $U/\Gamma=2.0$.} \label{figure_5a_5f}
\end{figure*}

\begin{figure*} [h]
\includegraphics[width=170mm]{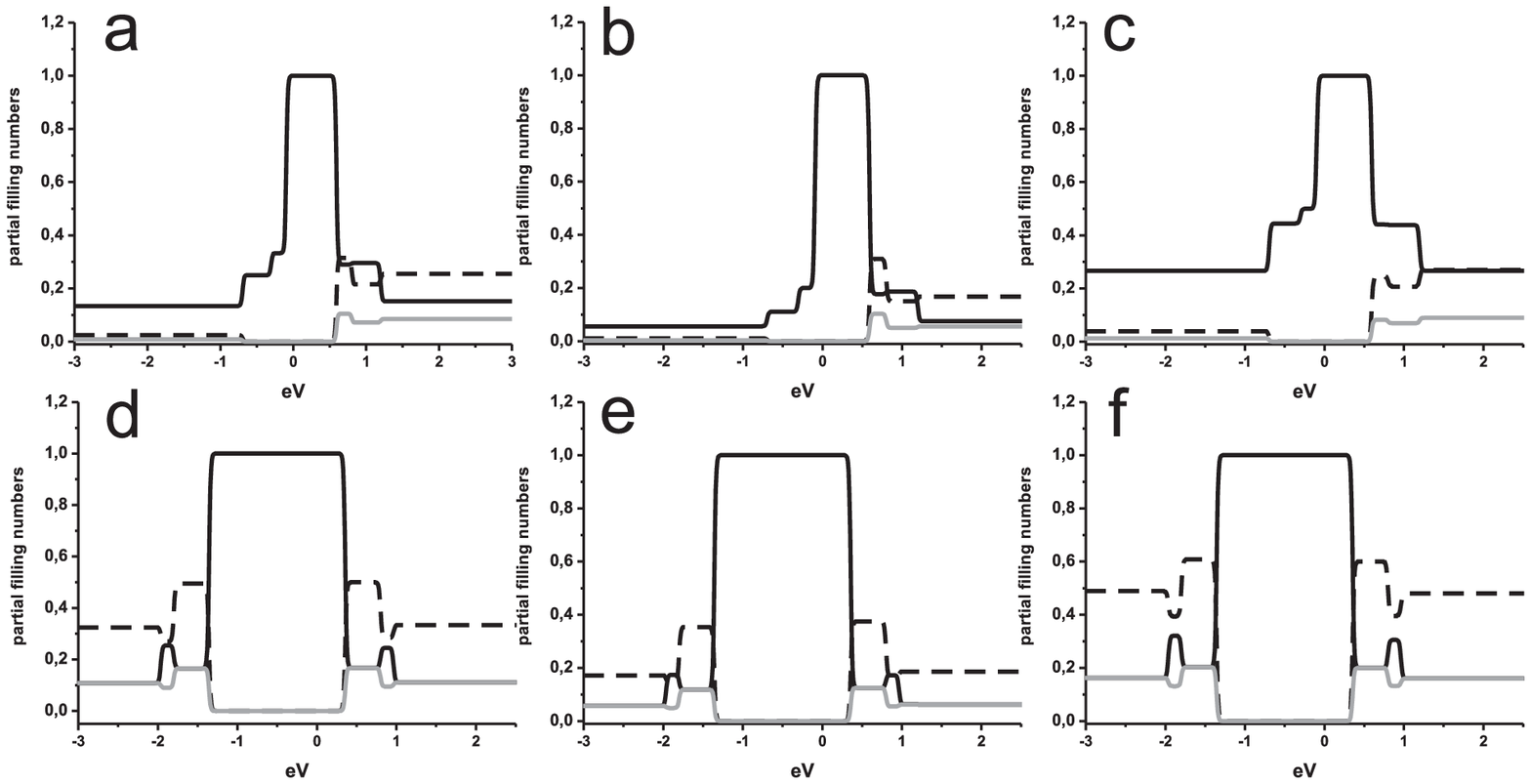}
\caption{The same dependencies as on the
Fig.\ref{figure_4a_4f};\ref{figure_5a_5f} but for both single
electron energy levels located below the sample Fermi level. Filling
numbers in the singlet state are shown by the black line, filling
numbers in the triplet state with the fixed projection of the spin
are shown by the grey line, full filling numbers in the triplet
state are shown by the black-dashed line. Parameters
$\varepsilon/\Gamma=-0.5$, $T/\Gamma=0.3$ are the same for all
figures. a).,d). $\Gamma_k=0.01$, $\Gamma_p=0.01$; b).,e).
$\Gamma_k=0.02$, $\Gamma_p=0.01$; c).,f). $\Gamma_k=0.01$,
$\Gamma_p=0.02$. a).-c). $U/\Gamma=0.5$; d).-f). $U/\Gamma=2.0$.}
\label{figure_6a_6f}
\end{figure*}

Another interesting feature which appears due to correlations is
multiple charge redistribution between the two electron states with
different spin configurations (singlet state and triplet state) as a
function of the applied bias voltage
(Fig.\ref{figure_4a_4f}-\ref{figure_6a_6f}).

In the case of both single electron energy levels situated above the
sample Fermi level (Fig.\ref{figure_4a_4f}) several ranges of
applied bias exist where the triplet state occupation exceeds the
singlet state occupation both for the strong ($U/\Gamma\gg
T/\Gamma$) and weak ($U/\Gamma\sim T/\Gamma$) Coulomb interaction.
In the case of strong Coulomb interaction localized charge is mostly
accumulated in the triplet state.

The occupation of the triplet state with the fixed value of the spin
projection $S_{Z}$ is lower than the the occupation of the singlet
state if Coulomb repulsion is weak ($U/\Gamma\sim T/\Gamma$)
(Fig.\ref{figure_4a_4f}a-c). In the case of strong Coulomb
interaction ($U/\Gamma\gg T/\Gamma$) four ranges exist where charge
is quite equally distributed between the singlet state and triplet
state with the fixed value of the spin projection
(Fig.\ref{figure_4a_4f}d,e,f) for different ratios between the
tunneling transfer rates.

Figure \ref{figure_5a_5f} demonstrates calculation results in the
case when one single electron energy level is located above and the
other - below the sample Fermi level. For the weak Coulomb
interaction ($U/\Gamma\sim T/\Gamma$)the charge is quite completely
localized in the singlet state both in the symmetric and asymmetric
tunneling contact (Fig.\ref{figure_5a_5f}a-c).

In the case of strong Coulomb interaction ($U/\Gamma\gg T/\Gamma$)
 charge in the system is also
mostly located on the singlet state for the wide range of applied
bias (Fig.\ref{figure_5a_5f}d-f), but for  some ranges of the
applied bias occupation of the triplet state exceeds occupation of
the singlet state. But for symmetric tunneling contact the triplet
state with the fixed spin projection is always less occupied than
the singlet state .

The situation when both single electron energy levels are positioned
below the sample Fermi level is demonstrated in
Fig.\ref{figure_6a_6f}. In the case of weak Coulomb interaction only
one range of the applied bias exists where occupation of the triplet
state is equal to or even exceeds the occupation of the singlet
state (Fig.\ref{figure_6a_6f}a-c). Occupation of the triplet state
with the fixed spin projection is always lower than the occupation
of the singlet state. The increasing of the Coulomb interaction
leads to the formation of several ranges of the applied bias where
triplet state filling numbers exceed singlet state filling numbers
(Fig.\ref{figure_6a_6f}d-f).

\section{Conclusion}

We investigated tunneling through the system of two interacting QDs
weakly coupled to the electrodes with Coulomb interaction between
localized electrons. In the considered system if electrons number
$N$ is changed
 due to the tunneling processes  the
modification of the energy spectrum is not reduced to the simple
adding of Coulomb interaction $U$ per electron. One, two, three or
four electrons can be localized in the coupled QDs , each state with
fixed total charge and spin projection has it's own energy.
Transitions between these states were analyzed in terms of
pseudo-particle operators with constraint on the possible physical
states of the system. Filling numbers of different multi-electron
states, total electron occupation of QDs and $I-V$ characteristics
were investigated for different single electron levels positions
relative to the sample Fermi level and various tunneling transfer
rates.

It was shown that total electron occupation demonstrates in some
cases significant decreasing with increasing of applied bias -
contrary to the situation with no correlations.

We revealed that for some parameter range, the system demonstrates
negative tunneling conductivity in certain ranges of the applied
bias voltage due to the Coulomb correlations. A negative tunneling
conductivity is well pronounced if both energy levels are located
above the Fermi level. When energy levels are located on the
opposite sites of the Fermi level or both of them are positioned
below the Fermi level negative tunneling conductivity was not
observed.

Coulomb correlations result in multiple charge redistribution
between the two-electron states with different spin configurations
(singlet and triplet states) with changing of applied bias voltage.
It was found that for particular range of the system parameters the
triplet-state occupation can exceed the singlet-state occupation
(the inverse occupation takes place).

So tunneling properties of correlated electron systems can be
correctly described only in terms of multi-electron states, which
allows to find some unexpected effects.

This work was partly supported by the RFBR and  Leading Scientific
School grants. The support from the Ministry of Science and
Education is also acknowledged.


\pagebreak

\end{document}